\documentclass[aps,a4paper,superscriptaddress,nofootinbib,tightenlines,floats]{revtex4}\usepackage{latexsym}
\usepackage{amssymb}
\usepackage[dvips]{graphicx}
\begin{document}
\renewcommand{\topfraction}{1}

\renewcommand{\=}{\!\!=\!\!}

\title{Fermion zero modes in $N\=2$ supervortices}
\author{A. Ach\'{u}carro} 
\affiliation{Lorentz Institute of
Theoretical Physics, University of Leiden, 2333 RA Leiden, The
Netherlands} \affiliation{Department of Theoretical Physics, UPV-EHU,
Bilbao, Spain} \author{A.C. Davis} \affiliation{DAMPT, Centre for
Mathematical Sciences, Cambridge University, U.K.}
\author{M. Pickles} \affiliation{DAMPT, Centre for Mathematical
Sciences, Cambridge University, U.K.}  
\author{J. Urrestilla}
\affiliation{Department of Theoretical Physics, UPV-EHU, Bilbao,
Spain} \date{\today}
\begin{abstract}
We study the fermionic zero modes of BPS semilocal magnetic vortices
in N=2 supersymmetric QED with a Fayet-Iliopoulos term and two matter
hypermultiplets of opposite charge. There is a one-parameter family of
vortices with arbitrarily wide magnetic cores.  Contrary to the
situation in pure Nielsen--Olesen vortices, new zero modes are found
which get their masses from Yukawa couplings to scalar fields that do
not wind and are non-zero at the core.  We clarify the relation
between fermion mass and zero modes. The new zero modes have opposite
chiralities and therefore do not affect the net counting (left minus
right) of zero modes coming from index theorems but manage to evade
other index theorems in the literature that count the total number (left
plus right) of zero modes in simpler systems.
\end{abstract}
\maketitle

\def\rd{{\rm{d}}}
\def\d{\partial}
\def\half{{1 \over 2}}
\def\quarter{{1 \over 4}}
\def\thetabar{\bar{\theta}}
\def\sech{{\rm sech}}
\def\tanh{{\rm tanh}}
\def\simleq{\; \raise0.3ex\hbox{$<$\kern-0.75em
      \raise-1.1ex\hbox{$\sim$}}\; }
\def\simgeq{\; \raise0.3ex\hbox{$>$\kern-0.75em
      \raise-1.1ex\hbox{$\sim$}}\; }  
\def\L{{\mathcal L}}
\def\t{\tilde}      
\def\bea{\begin{eqnarray}}
\def\eea{\end{eqnarray}}
\def\be{\begin{eqnarray}}
\def\ee{\end{eqnarray}}

\section{Introduction}

Stable magnetic flux tubes, or {\it strings}, occur in many high
energy models.  Fermionic zero modes on these strings can in some
cases render them superconducting, and dramatically change their
lifetime and interactions. In most of the examples that have been
studied in a cosmological context the strings are topological, yet the
effect of fermion zero modes on stable non--topological strings is
equally powerful and interesting.

In this paper we investigate the fermionic zero modes on magnetic
vortices in supersymmetric (SUSY) $N\=2$ QED with two hypermultiplets
of opposite charge.  This model was proposed as a toy model for
the low energy limit of type-II superstrings compactified on
Calabi--Yau manifolds, where the possibility of magnetic vortices was
first suggested in \cite{GMV96} (in the full low energy theory, there
are sixteen hypermultiplets charged under fifteen U(1) groups, that
is, fifteen copies of the model studied here).

In the absence of Fayet--Iliopoulos terms, it was proved in
\cite{ARH98} that the strings appearing in that model are unstable. In
\cite{ADPU01} the present authors studied the case where a (gauge)
symmetry breaking Fayet-Iliopoulos (FI) term was added. In this case
we found BPS string solutions. We also found that out of all the
possible vacua, only certain choices can give rise to static strings, due
to a {\it vacuum selection} effect \cite{PRTT96}. Even so, this effect
was still not enough to obtain stable strings, unlike in the case of a
comparable $N\=1$ supersymmetric model with two chiral multiplets \cite{ADPU01,PU02}.
The strings formed are semilocal strings \cite{VA91}, and in the BPS
limit (equal scalar and gauge masses) they are only neutrally stable
\cite{leese,H92}.

In the present work, a study of the possible fermionic zero modes
 arising in this model has been carried out, in order to see whether
 they stabilise the string by further augmenting the vacuum selection
 effect. If this were so, the semilocal strings could possibly turn
 into chiral cosmic strings \cite{CP99,DKPS00}, long-lived vortices in
 which the fermions move in one direction only. However we have found
 that this is not the case. The fermions
 respect the same vacuum selection effect as the bosons, meaning that
 the underlying string remains semilocal. Moreover, the
 fermions move in both directions along the string, and therefore they could mix and leave the string. 
It is known that the fermion back-reaction can modify the stability of the background string \cite{N95}.  We argue that this is not the case in the present context, because, unlike the systems studied in \cite{N95}, 
the Bogomol'nyi bound is protected by the remnant unbroken SUSY.
The ``topological charge'' is identical for all the vortices in the family, and we do not expect any member of the family to be singled out by the fermion back-reaction. 

Nevertheless, we have found that, due to the bosonic zero mode of the
semilocal string, one of the fermions is coupled to a boson which is
not zero at the core of the string, contrary to what might be expected
from related assertions in the literature. To our knowledge this is
the first time that the effect of charged, non--winding scalars has been
considered on fermion zero modes, and it is interesting to point out
that our model falls outside the scope of a number of index theorems
for vortex backgrounds in scalar--fermion systems.

In section II we present a brief review of the bosonic sector of this
 model; we then obtain fermionic zero modes by SUSY transformations of
 the bosonic background in section III. We check that these in fact
 satisfy the fermionic equations of motion in section IV, and derive
 the mass matrix. The results are then discussed in section V.

\section{Background}

We will investigate an $N\=2$ supersymmetric (SUSY) model consisting of
 two $N\=2$ hypermultiplets $(h_{ai},\psi_a,F_{ai})$, labelled by
 $a\=1,2$, with opposite charges $q_a\=\pm q$, coupled to an $N\=2$
 abelian vector multiplet $(A_\mu,M,N,\lambda^i,\vec{D})$ (in
 Wess-Zumino gauge), $i\=1,2$, plus a Fayet-Iliopoulos term $\vec{k}
 \cdot \vec{D}$, where $\vec{k}$ can be taken to be $(0,0,k)$ without
 loss of generality \cite{ADPU01}. This last term is responsible for
 breaking the $U(1)$ gauge symmetry, although it preserves the
 supersymmetry.

Using the conventions of \cite{sohnius}, the Lagrangian we are interested 
in can be written as

\bea 
\L&=&\half D^\mu h_a^{\, i} D_\mu h_{ai} + i \bar\psi_a
\gamma^\mu D_\mu \psi_a + \frac{i}{2}\bar\lambda_i\gamma^\mu\d_\mu
\lambda^i\nonumber\\ & &-\quarter F^{\mu \nu}F_{\mu \nu} + i \hat{q}_a
h_a^{\, i} \bar\lambda_ i \psi_a- i \hat{q}_a \bar\psi_a \lambda^i
h_{ai}\nonumber\\ & &-\half\left(H_1^{\, 1}-H_2^{\,
2}-\half\right)^2-\half\left(H_2^{\, 1} + H_1^{\, 2}\right)^2 -
\half\left(i H_2^{\, 1}-i H_1^{\, 2}\right)^2\,,
\label{lagrangian}
\eea 
where $\hat{q}_a=q_a/q$, $h_a^{\, i}=h^*_{aj}$, $H_j^{\,
i}=-(\hat{q}_a/2)h_a^{\, i}h_{aj}$, $D_\mu=\d_\mu + i \hat{q}_a A_\mu$
and $A_{\mu}$ is a U(1) gauge field.  Auxiliary fields have been
eliminated, all fields with zero vacuum expectation value (that
is, $M$ and $N$), have been set to zero and suitable rescalings have
been performed \cite{ADPU01}.

We can use a Bogomol'nyi argument to express the energy of 
the bosonic part of a static, straight vortex configuration as

\bea
E&=&\half \int \rd^2 x \left[\left|(D_1+iD_2)h_{11}\right|^2+\left|(D_1-iD_2)h_{12}\right|^2+\left|(D_1+iD_2)h_{21}\right|^2+
\left|(D_1-iD_2)h_{22}\right|^2+\right.\nonumber\\
& &\left.\left[B+(H_1^{\, 1}-H_2^{\, 2}+\half )\right]^2+(H_2^{\, 1}+H_1^{\, 2})^2+(i\,H_2^{\, 1}-i\,H_1^{\, 2})^2\right]\nonumber\\
& &-\half\int \rd^2 x B\,,
\eea
where $B\=\d_1 A_2-\d_2 A_1$, and the integral in the last term is the magnetic flux, which is quantised in units of 2$\pi$. The Bogomol'nyi equations can then be 
immediately read off

\bea
&(D_1+iD_2)h_{11}=0\,; \qquad \qquad (D_1-iD_2)h_{12}=0\,;&  \label{bogomolnyi}\\
&(D_1+iD_2)h_{21}=0\,; \qquad \qquad (D_1-iD_2)h_{22}=0\,;& \nonumber \\
&H_1^{\, 2}=H_2^{\, 1}=0\,; \qquad \qquad B+H_1^{\, 1}-H_2^{\,
2}+\half=0&\,.\nonumber
\eea
 
The only solutions to these equations have $h_{12}\=h_{21}\=0$
(this was dubbed a ``vacuum selection effect'' in \cite{ADPU01}) ; the
remaining equations are those of a semilocal string \cite{VA91,H92} in
$(h_{11},h^*_{22})$.

Up to global SU(2) transformations in the $(h_{11},h^*_{22})$ space, 
the unit winding, cylindrically symmetric semilocal string solution can be expressed as
\bea
h_{11}&=&f(r)e^{i\theta}\,;\nonumber\\
h_{22}&=&g(r)\,;\nonumber\\
A_\theta&=&a(r)\,. \label{semi}
\eea
with boundary conditions $f(0)\=a(0)\=g'(0)\=0$ and $f(\infty)\=1$, 
$g(\infty)\=0$ and $a(\infty)\=-1$. $g(r)$ is given by
\be
g(r)=\kappa \frac{f(r)}{r}\,.
\label{g}
\ee
where the constant $\kappa$ essentially measures the width of
the string, ranging from $\kappa\=0$ --the Nielsen-Olesen \cite{NO73}
string-- to $\kappa\=\infty$ --a $CP^1$ instanton \cite{H92}--. Note
the lack of winding in $h_{22}$ and furthermore that $h_{22}$ does not
necessarily have to be zero at $r\=0$. Note also that $B$ is always
negative with this choice of boundary condition, and its total flux
does not depend on $\kappa$.

\section{Fermionic zero modes}

We are now interested in studying the fermionic zero mode solutions to this
 system to see
 whether they influence this vacuum selection effect.
These solutions could be obtained by solving the explicit fermionic equations
 of motion,
but we can also use the SUSY transformation to get the zero modes directly 
\cite{DDT97}. This will give us static configurations in the plane 
perpendicular to the string, and we will subsequently introduce $t$
 and $z$ dependence on the solutions.

By zero mode solutions, we mean infinitesimal changes to the background 
configuration that preserve the
action (for static configurations, that amounts to leaving the energy
unchanged) and satisfy their equations of motion. We know that a SUSY
transformation of a given configuration leaves the energy
unchanged. Moreover, as we started with a static solution to the bosonic
equations of motion, the fermions produced by this transformation must
automatically satisfy their equations of motion.

The fermionic content of our system consists of two higgsinos (two Dirac fermions $\psi_1$ and 
$\psi_2$)
coming from the hypermultiplets, and two
gauginos (two symplectic Majorana fermions $\lambda^1$ and $\lambda^2$)
coming from the gauge vector multiplet. Recall that the symplectic Majorana spinors are
4-component $SU(2)$ covariant spinors defined from 2-component spinors
$\lambda_{\alpha i}$ and $\bar\lambda^i_{\dot\alpha}\equiv(\lambda_{\alpha i})^\dagger$ as

\be
\lambda^i\equiv\pmatrix{ -i \varepsilon^{ij}\lambda_{\alpha j}
\cr \bar\lambda^{\dot\alpha i} }\,;\qquad \bar\lambda_i=\left(\lambda_i^\alpha,
i\varepsilon_{ij}\bar\lambda^j_{\,\dot\alpha}\right)\,,
\label{sympl}
\ee
where $\varepsilon_{12}=\varepsilon^{12}=-\varepsilon_{\dot 1\dot 2}=-\varepsilon^{\dot 1\dot 2}=1$.

In $N\=1$ SUSY 
language, the gauginos can be thought of as consisting of one gaugino
coming from the $N\=1$ gauge multiplet and one higgsino coming 
from a neutral (with respect to the gauge multiplet) chiral multiplet.
The two SUSY generators ($\xi_{\alpha(1)}$, $\xi_{\alpha(2)}$) will be 
combined into two symplectic Majorana fermions $\epsilon^1$, $\epsilon^2$
 (\ref{sympl}).

Let us perform a SUSY transformation of the system in the 
background of the (bosonic) semilocal string obtained in the previous 
section. The bosonic fields do not transform. The higgsinos take the form

\bea
\delta\psi_{(a)}=-i \gamma^\mu D_\mu \epsilon^{(i)} h_{ai}\,,
\eea
while the gauginos may be written as

\bea
\delta\lambda^{(i)}=-\frac{i}{2}\sigma^{\mu\nu}\epsilon^{(i)}F_{\mu\nu}-i\epsilon^{(j)}
\,\vec{\tau}^{(i)}_{\,(j)}\cdot \vec{D}\,.
\eea

\noindent
Our conventions are \cite{sohnius}
\be
\gamma^\mu=\pmatrix{0 & \sigma^\mu \cr \bar\sigma^\mu & 0}\,, \quad \half i\left[\gamma^\mu,\gamma^\nu\right]=\sigma^{\mu\nu}\,,
\ee
where, $\sigma^\mu=(1,\mathbf{\sigma})$, $\bar\sigma^\mu=(1,-\mathbb{\sigma})$
and $\sigma^i$ are the Pauli matrices.

The higgsinos and the gauginos can be written more explicitly as 
\bea
\delta\psi_{(1)}&=&-\frac{i}{2}\left[\left(D_1-iD_2\right)h_{11}
\left(\gamma^1+i\gamma^2\right)\epsilon^{(1)}+\left(D_1+iD_2\right)h_{12}
\left(\gamma^1-i\gamma^2\right)\epsilon^{(2)}\right]\,;
\nonumber\\
\delta\psi_{(2)}&=&-\frac{i}{2}\left[\left(D_1+iD_2\right)h_{22}
\left(\gamma^1-i\gamma^2\right)\epsilon^{(2)}+\left(D_1-iD_2\right)h_{21}
\left(\gamma^1+i\gamma^2\right)\epsilon^{(1)}\right]\,;\nonumber\\
\delta\lambda^{(1)}&=&\gamma^1\gamma^2\epsilon^{(1)} B-
i \left(H_1^{\, 1}-H_2^{\, 2}+ \half \right)\epsilon^{(1)}\,;\nonumber\\
\delta\lambda^{(2)}&=&\gamma^1\gamma^2\epsilon^{(2)} B +
i \left(H_1^{\, 1}-H_2^{\, 2}+\half \right)\epsilon^{(2)}\,,
\eea
where we have used the Bogomol'nyi equations (\ref{bogomolnyi}).

We expect the strings to be $\half$BPS saturated \cite{WO78}, so let us try to obtain the broken and 
unbroken part of the SUSY
transformation, i.e., let us obtain the fermionic zero modes. 
In order to do so, 
we can use the following projectors

\be
P_{\pm}\equiv \half\left(1\pm i\gamma^1\gamma^2\right)\,,
\ee
which with our conventions, are given by
\bea
P_+\equiv{\rm diag}(1,0,1,0)\,,\qquad P_-\equiv{\rm diag}(0,1,0,1)\,.
\eea

These projectors, besides $P_\pm ^2= P_\pm ^{\ \dagger}= P_\pm$ and $P_\pm P_\mp=0$, have the following properties:

\bea
& &\gamma^1P_{\pm}=P_{\mp}\gamma^1\,;\nonumber\\
& &\gamma^2P_{\pm}=P_{\mp}\gamma^2\,;\nonumber\\
& &P_{\pm}\gamma^1=\pm i P_\pm \gamma^2\,.
\eea
$i\gamma^1\gamma^2$ is essentially a two-dimensional version of 
$\gamma^5$, acting in the plane perpendicular to the string.
Applying these projectors onto the fermions, we learn that

\bea
P_+\delta\psi_{(1)}&\equiv&\delta\psi_{(1)+}=-i\left(D_1-iD_2\right)h_{11}\gamma^1 P_-\epsilon^{(1)}
=-2iD_1 h_{11}\gamma^1P_-\epsilon^{(1)}\,;\nonumber\\
P_-\delta\psi_{(1)}&\equiv&\delta\psi_{(1)-}=-i\left(D_1+iD_2\right)h_{12}\gamma^1 P_+\epsilon^{(2)}
=0\,;\nonumber\\
P_+\delta\psi_{(2)}&\equiv&\delta\psi_{(2)+}=-i\left(D_1-iD_2\right)h_{21}\gamma^1 P_-\epsilon^{(1)}
=0\,;\nonumber\\
P_-\delta\psi_{(2)}&\equiv&\delta\psi_{(2)-}=-i\left(D_1+iD_2\right)h_{22}\gamma^1 P_+ \epsilon^{(2)}
=-2iD_1h_{22}\gamma^1P_+ \epsilon^{(2)}\,,
\eea

and

\bea
P_+\delta\lambda^{(1)}&\equiv&\delta\lambda^{(1)}_+=-i\left(B+H_1^{\, 1}-H_2^{\, 2}+1\right)
P_+\epsilon^{(1)}=0\,;\nonumber\\
P_-\delta\lambda^{(1)}&\equiv&\delta\lambda^{(1)}_-=i\left(B-H_1^{\, 1}+H_2^{\, 2}-1\right)
P_-\epsilon^{(1)}=2iBP_-\epsilon^{(1)}\,;\nonumber\\
P_+\delta\lambda^{(2)}&\equiv&\delta\lambda^{(2)}_+=-i\left(B-H_1^{\, 1}+H_2^{\, 2}-1\right)
P_+\epsilon^{(2)}=-2iBP_+\epsilon^{(2)}\,;\nonumber\\
P_-\delta\lambda^{(2)}&\equiv&\delta\lambda^{(2)}_-=i\left(B+H_1^{\, 1}-H_2^{\, 2}+1\right)
P_-\epsilon^{(2)}=0\,.
\eea

Note that $\delta\psi_{(1)-}$ and $\delta\psi_{(2)+}$ vanish on
BPS states due to the vacuum selection effect $h_{12}=h_{21}=0$.

It is clear that $P_-\epsilon^{(1)}$ and $P_+\epsilon^{(2)}$ generate
 the fermionic zero modes, whereas $P_+\epsilon^{(1)}$ and
 $P_-\epsilon^{(2)}$ are the generators of the unbroken SUSY. Note
 that there are fermionic zero modes of both chiralities, since N=2
 SUSY is non--chiral.

\section{Fermionic equations of motion}

In this section we investigate the structure of the fermionic
zero modes by analysing the equations of motion directly, without reference to supersymmetry. We use two-spinor notation and define

\be
\delta\psi_{(1)}\equiv\pmatrix{ \phi_{\alpha (1)} \cr
\bar{\chi}^{\dot{\alpha}}_{(1)}} \qquad\qquad
\delta\psi_{(2)}\equiv\pmatrix{ \phi_{\alpha (2)} \cr
\bar{\chi}^{\dot{\alpha}}_{(2)}} \qquad\qquad
\delta\lambda^{(1)}\equiv\pmatrix{ -i \Lambda_{\alpha(2)} \cr
\bar{\Lambda}^{\dot{\alpha}(1)} }\qquad\qquad
\delta\lambda^{(2)}\equiv\pmatrix{ i \Lambda_{\alpha (1)} \cr 
\bar{\Lambda}^{\dot{\alpha}(2)} }
\ee

The projector operators can be defined in two spinor notation as 
\be
\sigma_+=\pmatrix{1&0\cr 0&0}\,;\qquad\qquad\sigma_-=\pmatrix{0&0 \cr 0&1}\,.
\label{twopro}
\ee

In this notation the zero modes that we found previously are
\bea
\phi_{\alpha (1)}&=& -2i D_1 h_{11} \sigma^1 \sigma_- \bar\xi^{\dot\alpha(1)}\,;\nonumber\\
\bar{\chi}^{\dot{\alpha}(1)}&=&-2 D_1 h_{11}\bar\sigma^1\sigma_-\xi_{\alpha(2)}\,;\nonumber\\
\phi_{\alpha (2)}&=&
-2 iD_1 h_{22} \sigma^1 \sigma_+ \bar\xi^{\dot\alpha(2)}\,;\nonumber\\
\bar{\chi}^{\dot{\alpha}(2)}&=&2  D_1 h_{22} \bar\sigma^1 \sigma_+ \xi_{\alpha(1)}\,;\nonumber\\
\Lambda_{\alpha(1)}&=&-2iB\sigma_+\xi_{\alpha(1)}\,;\nonumber\\
\Lambda_{\alpha (2)}&=&2iB\sigma_-\bar\xi_{\alpha(2)}\,,
\label{feomdef}
\eea
where $h_{11}, \ h_{22}$ and $B$ satisfy  (\ref{bogomolnyi}).

Consider the fermionic equations of motion derived from
(\ref{lagrangian}) in the bosonic background given by a solution to
(\ref{bogomolnyi}). Recall that $h_{12} = h_{21} = 0$, and all other
fields are independent of $t$ and $z$:

\bea
\left(\bar{\sigma}^\mu\right) D_{\mu}\phi_{(1)}
-\bar{\Lambda}^{(1)} h_{11}&=&0\,;\nonumber\\
\left(\sigma^\mu\right) D_\mu\bar{\chi}_{(1)}+i
\Lambda_{(2)} h_{11}&=&0\,;\nonumber\\
\left(\bar{\sigma}^\mu\right) D_{\mu}\phi_{(2)}
+\bar{\Lambda}^{(2)} h_{22}&=&0\,;\nonumber\\
\left(\sigma^\mu\right) D_\mu\bar{\chi}_{(2)}+i
\Lambda_{(1)} h_{22}&=&0\,;\nonumber\\
\left(\sigma^\mu\right) \d_{\mu}\bar{\Lambda}^{(1)}
+h^*_{11}\phi_{(1)}+i\chi_{(2)} h_{22}&=&0\,;\nonumber\\
\left(\sigma^\mu\right) \d_{\mu}\bar{\Lambda}^{(2)}
-h^*_{22}\phi_{(2)}+i\chi_{(1)} h_{11}&=&0\,.
\label{feom}
\eea

It can be checked that the static, $z$--independent
configurations (\ref{feomdef}) satisfy these equations. Including the $(t,z)$
dependence into (\ref{feom}) we learn that three
of the fermions are functions of $(t-z)$ and therefore move in the
positive $z$ direction, while the other three move in the negative $z$
direction: 

\bea
\phi_{\alpha (1)},\, \chi_{\alpha (2)},\, \Lambda_{\alpha (1)} &\to& t+z\,;
\nonumber\\
\phi_{\alpha (2)},\, \chi_{\alpha (1)},\, \Lambda_{\alpha (2)} &\to& t-z\,.
\eea

The fermions move in opposite directions, as expected, since $N\=2$ 
SUSY is intrinsically non-chiral and it has not been broken in this model.
We are interested in the $(r, \theta)$ dependence of the zero modes.

The usual Nielsen-Olesen string is one of the possible configurations
in the family of semilocal strings, and it corresponds to the
narrowest string. One can characterise this string as having
$h_{22}\=0$, annihilating two out of the six Weyl fermions
$\phi_{(2)}\=\bar{\chi}_{(2)}\=0$ \cite{hou}.  Moreover, if we remove
one SUSY generator out of the two, we recover the Nielsen-Olesen
string, but with chiral fermions \cite{DDT97} moving in the same
direction. This is a good check of our results.

\begin{figure}[!htb]
\begin{center}
\includegraphics[width=7cm]{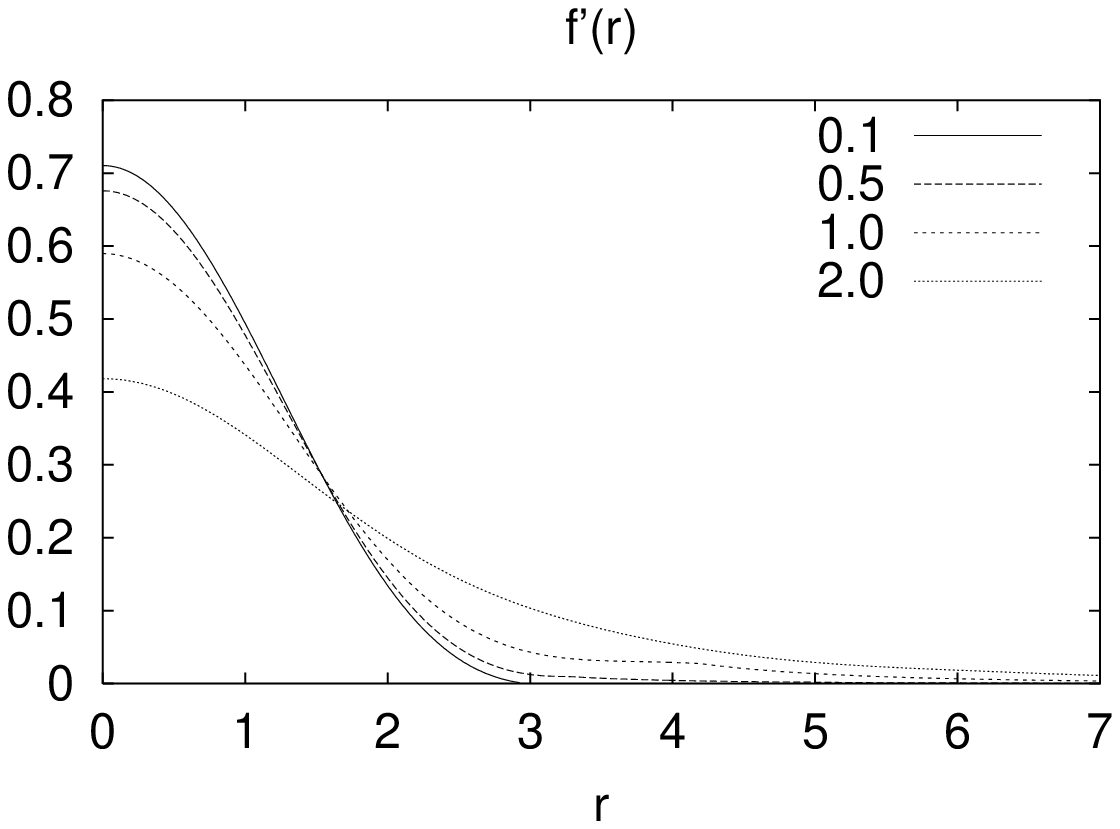}
\hspace{1cm}
\includegraphics[width=7cm]{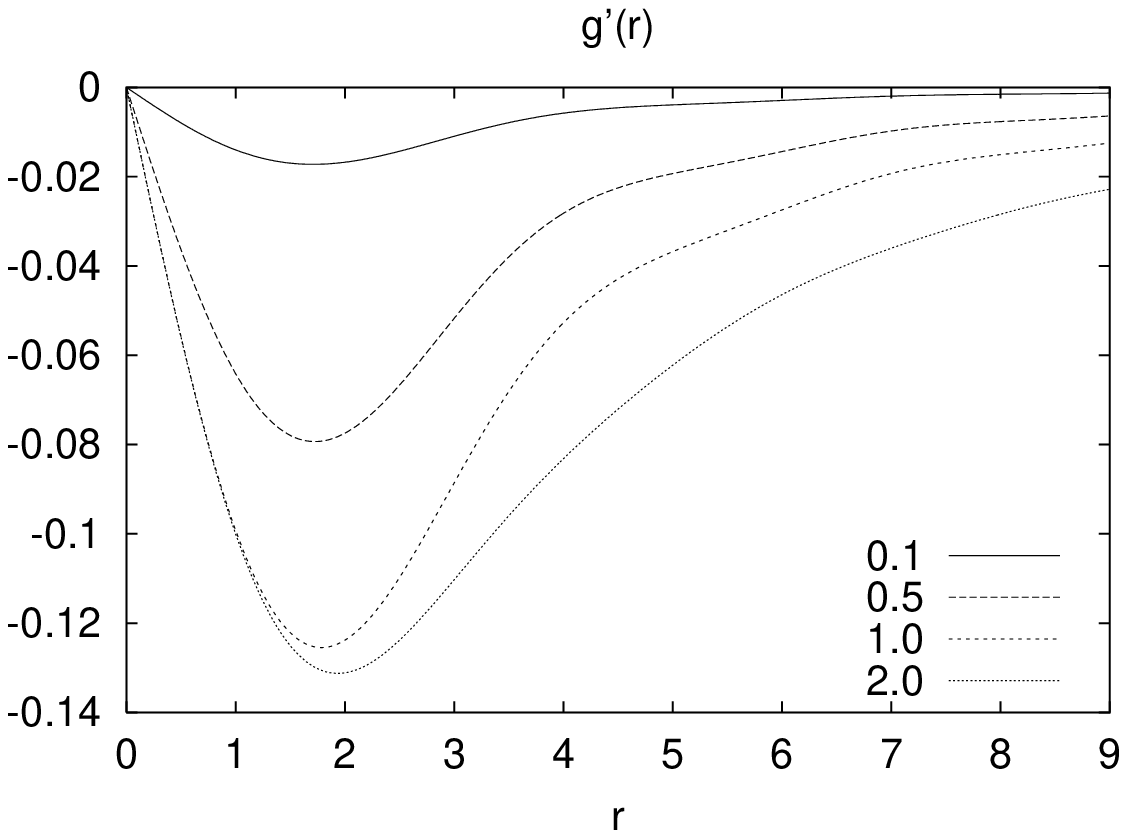}\\
\caption[fig]{\label{fig} Profiles of the derivatives of the functions $f(r)$ and $g(r)$ given by equations (\ref{semi}) and 
(\ref{g}), for
different values of the parameter $\kappa$.}
\end{center}
\end{figure}

The situation is different when $h_{22}\!\!\ne\!\!0$. The fermions
$\phi_{(2)}$ and $\chi_{(2)}$ are coupled to a field which is not zero
at $r\=0$, which might seem surprising.  In terms of the general
string ansatz (\ref{semi}), these fermions may be expressed in the
form 
\bea
\phi_{(2)}&=&-2i\kappa\d_r\left(\frac{f(r)}{r}\right)e^{i\theta}\pmatrix{0\cr\bar\xi^{\dot 1 (2)}}\,;\nonumber\\ 
\chi_{(2)}&=&-2\kappa\d_r\left(\frac{f(r)}{r}\right)e^{-i\theta}\pmatrix{\bar\xi^{\dot2 (1)}\cr 0}\,,
\eea
and, as can be seen from figure~\ref{fig}, they tend to $0$ at $r\=0$. These
two fermions are the only ones that wind.

In order to 
analyse the fermion zero modes, we convert the Dirac equations
(\ref{feom}) into second order equations by acting with the operators
$\sigma\cdot D$ and  $\bar\sigma\cdot D$:

\bea
& &(\Box+\sigma^3 B+|h_{11}|^2)\phi_{(1)}+ih_{11}h_{22}\chi_{(2)}-(\sigma\cdot D)h_{11}\bar\Lambda^{(1)}=0\,;\nonumber\\
& &(\Box+\sigma^3 B+|h_{22}|^2)\chi_{(2)}-ih_{11}^*h_{22}^*\phi_{(1)}+
i(\sigma\cdot D)h_{22}^*\bar\Lambda^{(1)}=0\,;\nonumber\\
& &(\Box+|h_{11}|^2+|h_{22}|^2)\bar\Lambda^{(1)}+(\bar\sigma\cdot D)h_{11}^*\phi_{(1)}+i(\bar\sigma \cdot D)h_{22}\chi_{(2)}=0\,;\nonumber\\
& &(\Box-\sigma^3 B+|h_{22}|^2)\phi_{(2)}-ih_{11}h_{22}\chi_{(1)}+(\sigma\cdot D)h_{22}\bar\Lambda^{(2)}=0\,;\nonumber\\
& &(\Box-\sigma^3 B+|h_{11}|^2)\chi_{(1)}+ih_{11}^*h_{22}^*\phi_{(2)}+
i(\sigma\cdot D)h_{11}^*\bar\Lambda^{(2)}=0\,;\nonumber\\
& &(\Box+|h_{11}|^2+|h_{22}|^2)\bar\Lambda^{(2)}-(\bar\sigma\cdot D)h_{22}^*\phi_{(2)}+i(\bar\sigma \cdot D)h_{11}\chi_{(1)}=0\,,
\label{box1}
\eea
where $\sigma\cdot D\=\sigma^1D_1+\sigma^2D_2$ and 
$\bar\sigma^{1,2}=-\sigma^{1,2}$.

We will use two dimensional projectors also in this case;
and we define the projected two-spinors $\Psi_\pm$ by $\Psi_\pm \equiv \sigma_\pm \Psi$, 
for any two-spinor $\Psi$, where $\sigma_\pm$ are given in equation (\ref{twopro}).

Half of the equations (\ref{box1}) correspond to the fermions coming
from the unbroken SUSY symmetry:
\bea
& &(\Box-B+|h_{11}|^2)\phi_{(1)-}+ih_{11}h_{22}\chi_{(2)-}=0\,;\nonumber\\
& &(\Box-B+|h_{22}|^2)\chi_{(2)-}-ih_{11}^*h_{22}^*\phi_{(1)-}=0\,;\nonumber\\
& &(\Box+|h_{11}|^2+|h_{22}|^2)\bar\Lambda^{(1)}_+=0\,;\nonumber\\
& &(\Box-B+|h_{22}|^2)\phi_{(2)+}-ih_{11}h_{22}\chi_{(1)+}=0\,;\nonumber\\
& &(\Box-B+|h_{11}|^2)\chi_{(1)+}+ih_{11}^*h_{22}^*\phi_{(2)+}=0\,;\nonumber\\
& &(\Box+|h_{11}|^2+|h_{22}|^2)\bar\Lambda^{(2)}_-=0\,.
\label{eq1}
\eea

We can think of these equations as giving position--dependent {\it
masses} (squared) for the fermions, in the sense that after
diagonalisation, all six equations are of the form $\Box \Psi + {\it
M^2} \Psi = 0$, where

\be {\it M^2}={\rm diag} ( -B +(|h_{11}|^2 + |h_{22}|^2), -B, |h_{11}|^2 +
|h_{22}|^2, -B +(|h_{11}|^2 + |h_{22}|^2), -B, |h_{11}|^2 +
|h_{22}|^2)\,. 
\label{m}
\ee

Note that the matrix $M^2$ is not the same as the fermion mass matrix
squared, due to the presence of derivative terms (in particular $M^2$
involves the magnetic field strength and is gauge invariant).  Both
matrices of course agree on constant bosonic backgrounds. As a simple
check, far from the core we recover the fermion masses;
indeed, $B \to 0$, $h_{11} \to 1$ and $h_{22} \to 0$ as $r \to
\infty$, and the diagonal terms become $1,0,1$, the correct masses for
one higgsino, one goldstino and one gaugino of each chirality.

Recall that we are using signature $(+,-,-,-)$. Since $\Box = -
\nabla^2$, the laplacian in the plane perpendicular to the string,
(\ref{eq1}) is a set of coupled time--independent Schr\"odinger
equations in $(r, \theta)$ and we are looking for the zero energy
states. Obviously, if the ``potential'' is everywhere positive there
are no zero energy eigenstates. The functions in ${\it M^2}$ (\ref{m})
are all greater than zero everywhere, and thus there are no non-zero
normalizable solutions to (\ref{eq1}), as can be shown by simply
multiplying by $\Psi^*$ and integrating by parts.

Thus {\it all} of these fermionic components are automatically zero
because of the equations of motion, and this agrees with the form of
the solution obtained from SUSY transformations (\ref{feom}).

The other half of the equations reads 
\bea &
&(\Box+B+|h_{11}|^2)\phi_{(1)+}+ih_{11}h_{22}\chi_{(2)+}-2\sigma^2
\bar\Lambda^{(1)}_-D_2 h_{11}=0\,;\nonumber\\ &
&(\Box+B+|h_{22}|^2)\chi_{(2)+}-ih_{11}^*h_{22}^*\phi_{(1)+}+2i\sigma^2
\bar\Lambda^{(1)}_- D_2 h_{22}^*=0\,;\nonumber\\ &
&(\Box+|h_{11}|^2+|h_{22}|^2)\sigma^2 \bar\Lambda^{(1)}_- -
2\phi_{(1)+} D_2 h_{11}^* - 2i \chi_{(2)+} D_2 h_{22} =0\,;\nonumber\\
& &(\Box+B+|h_{22}|^2)\phi_{(2)-}-ih_{11}h_{22}\chi_{(1)-}+2 \sigma^2
\bar\Lambda^{(2)}_+ D_2 h_{22}=0\,;\nonumber\\ &
&(\Box+B+|h_{11}|^2)\chi_{(1)-}+ih_{11}^*h_{22}^*\phi_{(2)-}+2i\sigma^2
\bar\Lambda^{(2)}_+ D_2 h_{11}^* =0\,;\nonumber\\ &
&(\Box+|h_{11}|^2+|h_{22}|^2)\sigma^2 \bar\Lambda^{(2)}_+ +
2\phi_{(2)-} D_2 h_{22}^* -2i \chi_{(1)-} D_2 h_{11} =0\,, 
\eea 
where, again, the Bogomol'nyi equations (\ref{bogomolnyi}) have been
used extensively.

These  may be written as two sets of three coupled equations\footnote{Note that each set of equations correspond to fermions of one given chirality and
coming from one of the supersymmetry generators} 
 and we
again seek to diagonalize the resulting mass matrices.

\be
\Box \pmatrix{ \phi_{(1)+} \cr \chi_{(2)+} \cr \sigma^2 \bar\Lambda^{(1)}_- 
} + \pmatrix{ 
B + |h_{11}|^2 & i h_{11} h_{22} & -2 D_2 h_{11} \cr
-i h_{11}^* h_{22}^* & B+|h_{22}|^2  & 2i D_2 h_{22}^* \cr
-2 D_2 h_{11}^* & -2i D_2 h_{22} & |h_{11}|^2 + |h_{22}|^2}
\pmatrix{ \phi_{(1)+} \cr \chi_{(2)+} \cr \sigma^2 \bar\Lambda^{(1)}_- } = 0
\ee
\be
\Box \pmatrix{ \phi_{(2)-} \cr \chi_{(1)-} \cr \sigma^2\bar\Lambda^{(2)_+ } }+  \pmatrix{
 B+|h_{22}|^2 & -i h_{11} h_{22} & 2 D_2 h_{22} \cr
 i h_{11}^* h_{22}^* & B+|h_{11}|^2 & 2i D_2 h_{11}^* \cr
 2 D_2 h_{22}^* & -2i D_2 h_{11} & |h_{11}|^2 + |h_{22}|^2}
\pmatrix{ \phi_{(2)-} \cr \chi_{(1)-} \cr \sigma^2\bar\Lambda^{(2)_+ } } = 0
\ee

For simplicity we just consider the first set of three equations. In
the case of the Nielsen-Olesen string member of the semilocal family ,
for which $\kappa\=0$, $h_{22}\=0$, the spinors $\chi_{(2)+}$ and
$\phi_{(2)-}$ do not couple to any spinors, and have mass squared B,
which is negative in the core of the string, and zero at infinity.
After diagonalisation, the other mass squared terms are \be \half (B +
2 |h_{11}|^2 \pm \sqrt{ B^2 + 16 D_2 h_{11} D_2 h_{11}^* } ) \ee which
correspond to two (r-dependent) linear combinations of $\phi_{(1)+}$
and $\sigma^2 \bar\Lambda^{(1)}_-$.  At $r\=0$, the signs of these
masses are $+,-$ respectively. At infinity these masses are 1, 1, and
the mass states are simply the uncombined spinors.  We can immediately
see by the same reasoning as before that one combination of fermions
is zero everywhere, since its mass squared is positive everywhere. The
other one is a combination of the higgsino and the gaugino.

For the general case, $\kappa \neq 0$, we have to diagonalize two sets
 of $3\times 3$ matrices.  The signs of the ``eigenvalues'' of the
 $M^2$  matrices are $+,-,-$ at $r\=0$ and $+,0,+$ at infinity, the
 ``eigenvectors'' being a combination of the three fermions. Once
 again, the ``eigenvector'' whose mass--squared function is always
 positive is zero, and thus we are left with two non-zero
 ``eigenvectors''.

In all cases, the fermions at infinity have
masses $1,0,1,1,0,1$, which agree with the masses of the fields
$h_{11},h_{12},A_\mu,h_{22},h_{21},M+iN$, as should be the case since
supersymmetry is unbroken there.

\section{Discussion}

We have studied the fermion zero modes arising in SUSY $N\=2$ QED with two hypermultiplets and a FI term. It is known that in such a model
there is a vacuum selection effect in the bosonic sector, and only some of all the possible vacua form a family of neutrally stable vortices \cite{ADPU01}. We showed that the fermion zero modes do not improve the stability of 
 the vortices. This is due to the fact that fermions obey the same vacuum selection effect as the vortices. On the other hand, as the SUSY is $N\=2$, fermions with both chiralities are present in the vortex.

We do not expect the stability to be altered by the back-reaction of the fermions either. A recent detailed calculation of quantum corrections for SUSY kinks shows that the Bogomol'nyi bound is preserved \cite{G}), even if the mass receives corrections.  

The model analysed in this paper contains a family of vortices that have the same topological charge, and all saturate the Bogomol'nyi bound. SUSY is half broken for all members of the family, and the unbroken SUSY will preserve the Bogomol'nyi bound in all cases, since the multiplets are shortened.
Thus the BPS condition holds for all members of the family, and as they all have the same topological charge, the energy of all of them is also the same. Thus, no member of the family will be singled out by fermionic back-reaction.

The fermions in this system are also interesting because some are 
coupled to a scalar field which is not zero at the core of the vortex.
It is sometimes stated in the literature that the reason why 
vortices support fermion zero modes in their core is because the
fermion masses, which come from coupling to the scalars, are zero
there. We have shown here that this heuristic argument is incorrect by 
calculating explicitly the zero modes that couple to the non--zero
scalar at the core ($h_{22}$, if the string is in $h_{11}$).

It is interesting to try to relate the zero modes derived from
supersymmetry to the zero modes we see in the equations of
motion. The two fermionic zero modes obtained directly from
supersymmetry are related to translational zero modes in the bosonic
sector.

We can see from the second order equations of motion that in fact
there are two more fermionic zero modes: 
In the bosonic sector, there are zero modes corresponding to changes
in the parameter $\kappa$ of the semilocal string, expanding the
string core. This bosonic zero mode corresponds to the fermionic zero
modes that have negative mass at zero, and which are massless at
infinity. One way to see this is to take the case where there is just
one hypermultiplet $(h_i,\psi,F_i)$. The string formed would then be
an ordinary Nielsen-Olesen string, which would not possess the bosonic
zero mode. Furthermore, the spinors $\phi_2$ and $\chi_2$ would
disappear, and the remaining eigenstates would have signs $+,-,+,-$
for their masses squared at zero and $+,+,+,+$ at infinity, so that
only two of the eigenstates are non-zero. When we re-introduce the
second hypermultiplet, the extra fermionic zero modes correspond to
the extra bosonic zero mode associated with $h_{22}$.

It is also possible to consider supersymmetric transformations of the
bosonic zero mode to give the fermionic zero mode. We perturb the
field $h_{22}$, adjusting the other fields so that we retain the
Bogomol'nyi equations (\ref{bogomolnyi}), and hence keep the same
energy.  In the limit $\kappa=0$, the semilocal string has $h_{22}=0$,
and an infinitesimal perturbation of $h_{22}$ of the form $\delta
h_{22}= \alpha h_{11\,(background)}/r$ does not modify the other
fields. Hence, a supersymmetry transformation of this zero mode gives
fermionic zero modes in $\phi_{(2)}$ and $\chi_{(2)}$ only.

At infinity the bosonic zero mode corresponding to changing the string
width is a goldstone boson, and this agrees with what we have
discovered from the fermionic equations of motion: that the
eigenstates that have mass zero at infinity are pure $\phi_{(2)}$ or
$\chi_{(2)}$ in the $\kappa\=0$ case.

For a general semilocal string $(\kappa\neq 0)$, all the bosonic
fields will be affected by a perturbation in the string width, and so
the corresponding fermionic zero modes are combinations of each set of
three spinors.

We note that the index theorem of Davis, Davis and Perkins
\cite{DDP97} does not apply in the case of the semilocal string due to
the fact that one of the fields doesn't wind in the core -- an
assumption made in the derivation of the index theorem. Hence we can't
use it to ascertain the number of zero modes in this case. In the case
where the index theorem does apply, namely the Nielsen--Olesen string
(i.e. one hypermultiplet), our results agree with the index
theorem. For the semilocal string there are two extra zero modes
corresponding to the change of the string width; the fermion zero
modes are those fermions corresponding to the bosonic zero mode. This
has been constructed explicitly in the Nielsen--Olesen limit of the
semilocal string. The model also falls outside the scope of the index
theorem of Ganoulis and Lazarides\cite{GL88}, because $h_{22}$ is
charged under $U(1)$ but goes to zero at infinity.  Our results agree
with Weinberg's index theorem\cite{W81} (see also \cite{JR81}), since
the new zero modes have opposite chiralities and therefore do not
affect the counting of net (left minus right) fermion zero modes. This
is in agreement with the fact that the magnetic flux measured at
infinity is the same for all semilocal strings, including the
Nielsen--Olesen string.

\section*{Acknowledgements}

We thank F. Freire, T. Vachaspati and P. van Baal for very useful
discussions.  This work was partially supported by the ESF COSLAB
programme (M.P.) and by grants AEN99-0315, FPA 2002-02037 and 9/UPV00172.310-14497/2002 (A.A., J.U.). J. Urrestilla and M. Pickles thank the University of Leiden for their hospitality.


\begin{thebibliography}{99}

\bibitem{GMV96} B.R.~Greene, D.R.~Morrison and C.~Vafa,
Nucl.~Phys.~{\bf B481}, 513 (1996)
\bibitem{ARH98} A.~Ach\'ucarro, M.~de~Roo, L.~Huisoon,
Phys.~Lett.~{\bf B424}, 288 (1998)
\bibitem{ADPU01} A.~Ach\'ucarro, A.C.~Davis, M.~Pickles,
J.~Urrestilla, Phys.~Rev.~{\bf D66}, 105013 (2002)
\bibitem{PRTT96}A.~A.~Penin, V.~A.~Rubakov, P.~G.~Tinyakov, S.~V.~Troi
tsky, Phys.~Lett.~{\bf B389}, 13 (1996)
\bibitem{PU02} M.~Pickles, J.~Urrestilla, JHEP {\bf 0301}, 052 (2003)

\bibitem{VA91} T.~Vachaspati and A.~Ach\'ucarro, Phys.~Rev.~{\bf D44}, 3067 (1991)
\bibitem{H92} M.~Hindmarsh, Phys.~Rev.~Lett.~{\bf 68}, 1263 (1992)
\bibitem{leese} R.A.~Leese, Phys.~Rev.~{\bf D46}, 4677 (1992)
\bibitem{CP99} B.~Carter, P.~Peter, Phys.~Lett.~{\bf B466}, 41 (1999)
\bibitem{DKPS00} A.C.~Davis, T.W.B.~Kibble, M.~Pickles, D.A.~Steer,
Phys.~Rev.~{\bf D62}, 083516 (2000)
\bibitem{N95} S.G.~Naculich, Phys.~Rev.~Lett.~{\bf 75}, 998 (1995); H.~Liu, T.~Vachaspati, Nucl.~Phys.~{\bf B470}, 176 (1996); M.~Groves, W.B.~Perkins, Nucl.~Phys.~{\bf B573}, 449 (2000)
\bibitem{sohnius} M.F.~Sohnius, Phys.~Rept.~{\bf 128}, 39 (1985)
\bibitem{NO73} H.~Nielsen, P.~Olesen, Nucl. Phys. {\bf B61}, 45
(1973)
\bibitem{DDT97} S.C.~Davis, A.C.~Davis, M.~Trodden, Phys.~Lett.~{\bf B405}, 257 (1997)
\bibitem{WO78} E.~Witten, D.I.~Olive, Phys.~Lett.~{\bf B78}, 97 (1978) 
\bibitem{hou} X.~Hou, Phys.~Rev.~{\bf D63}, 045015 (2001)
\bibitem{G} A.~Rebhan, P.~van Nieuwenhuizen, R.~Wimmer, Nucl.~Phys.~{\bf B648}, 174 (2003); A.S.~Goldhaber, A.~Rebhan, P.~van Nieuwenhuizen, R.~Wimmer ``Quantum corrections to the mass and central charge of solitons in $1+1$ dimensions'', based on a talk given by P. van Nieuwenhuizen at the 3rd Sakharov Conference On Physics, June 2002, Moscow; {\tt hep-th/0211087}
\bibitem{DDP97} S.C.~Davis, A.C.~Davis, W.B.~Perkins, Phys.~Lett.~{\bf
B408}, 81 (1997)
\bibitem{GL88} N.~Ganoulis, G.~Lazarides, Phys. ~Rev. ~{\bf D38}, 547  (1988)  
\bibitem{W81} E.J.~Weinberg, Phys. ~Rev. ~{\bf D24}, 2669  (1981)  
\bibitem{JR81} R.~Jackiw, P.~Rossi, Nucl.~Phys {\bf B190} 681 (1981).
\end{thebibliography}
\end{document}